\documentclass{ws-ijgmmp}

\usepackage{graphicx}  
\usepackage{latexsym}
\usepackage{amssymb}
\usepackage{amsmath}

\def\hook{\,\mathop{\breve{\,}}\nolimits\,}
\def\lefthook{\hook\kern-1.5pt}

\def\sot#1#2{\oalign{$#1$\crcr\hidewidth\vbox to.2ex{\hbox{$_{#2}$}
 \vss}\hidewidth}}

\def\leftcontract{\mathop{\hbox{\vrule height0.5pt width6pt \vrule width0.5pt
   height6pt}}}

\newcommand{\beq}{\begin{equation}}
\newcommand{\eeq}{\end{equation}}
\newcommand{\bea}{\begin{eqnarray}}
\newcommand{\eea}{\end{eqnarray}}
\def\eqref#1{(\ref{#1})}

\begin{document}

\markboth{D. Bini, A. Geralico, R.T. Jantzen}
{Wedging  spacetime principal null directions }

%
\catchline{}{}{}{}{}
%

\title{\uppercase{ 
Wedging spacetime principal null directions}
}

\author{\footnotesize DONATO BINI}

\address{
Istituto per le Applicazioni del Calcolo ``M. Picone,'' CNR, I--00185 Rome, Italy\\
INFN - Sezione di Roma III, Rome, Italy\\
\email{donato.bini@gmail.com}
}

\author{\footnotesize ANDREA GERALICO}

\address{
Istituto per le Applicazioni del Calcolo ``M. Picone,'' CNR, I--00185 Rome, Italy\\
\email{andrea.geralico@gmail.com}
}

\author{\footnotesize ROBERT T. JANTZEN}

\address{
Department of Mathematics and Statistics, Villanova University, Villanova, PA 19085, USA\\
\email{robert.jantzen@villanova.edu}
}

\maketitle

\begin{history}
\received{(Day Month Year)}
\revised{(Day Month Year)}
\end{history}

\begin{abstract}
Taking wedge products of the $p$ distinct principal null directions associated with the eigen-bivectors of the Weyl tensor associated with the Petrov classification, when linearly independent, one is able to express them in terms of the eigenvalues governing this decomposition. 
We study here algebraic and differential properties of such $p$-forms by completing previous geometrical results concerning type I spacetimes and extending that analysis to algebraically special spacetimes with at least 2 distinct principal null directions.
A number of vacuum and nonvacuum spacetimes are examined to illustrate the general treatment.
\end{abstract}


\section{Introduction}

The Petrov classification \cite{Stephani:2003tm} determines the algebraic type of the Weyl curvature tensor at a given point of a 4-dimensional spacetime by solving an eigenvalue problem. The corresponding eigen-bivectors are associated with certain real null vectors, which are called the ``principal null directions" (PNDs) of the considered spacetime.
In all algebraically special cases the dimension of the span of the set of PNDs equals the number of distinct (nonproportional)  such vectors due to the peculiar nature of null vectors in 4-dimensions: 3 (Type II), 2 (Type D, III), 1 (Type N) and 0 (Type O).
In the algebraically general case (Type I) the number of distinct eigenvalues and PNDs are both 4, but the span of the PNDs may have either dimension 3 or 4 \cite{Trumper,Penrose:1986ca,ArMcIn1994}. We consider only the latter case here, and the remaining cases with $p>1$ distinct linearly independent PNDs. 

In a previous article \cite{Bini:2023} we provided a geometrical criterion to distinguish the maximally spanning (dimension 4) and nonmaximally spanning (dimension 3) type I cases based on the nonvanishing or vanishing of the wedge product of their four distinct PNDs, respectively, focusing on its {\it on/off} behavior only.
Here, we investigate the geometrical properties of the associated 4-forms in the maximally spanning case together with their algebraic and differential structure, extending our analysis to the remaining nontrivial lower dimensional cases with multiple distinct PNDs, so the number of linearly independent PNDs is $p=2,3,4$.

Understanding the causal structure of tensor fields in any given spacetime, reflected in their algebraic invariants, is facilitated by decomposing them into parts associated with ``space plus time" projections through the introduction of a future pointing unit timelike vector $u$ picking out a temporal direction and interpretable as the 4-velocity of an observer \cite{Jantzen:1992rg}.
For $p$-forms this means decomposing them into electric and magnetic parts. In particular, every $p$-form determines a $p$-dimensional subspace of the vector space which contains some combinations of spatial, temporal and null directions. For some values of $p$, the eigendirections of such a $p$-form or those its spacetime dual, or those of its electric and magnetic parts, can offer a more efficient packaging of the associated information.

Null vectors cannot be normalized in the traditional way and their overall scale is irrelevant in the Petrov classification scheme where only their direction has significance. However, for a given observer 4-velocity $u$ in spacetime, this scale freedom can be fixed by requiring a unit timelike future-pointing projection of the null vector, thus representing it in the form $k=u+\hat\nu$.  Then its orthogonal spatial part $\hat\nu$ in the local rest space of that observer is the unit spatial velocity of the corresponding light ray aligned with the null vector, namely its direction in that local rest space. One can then express the suitably rescaled PNDs and their wedge products and various contraction scalars with themselves and the Weyl tensor explicitly first in terms of the  set of associated light ray directions.
When the observer $u$ is the timelike member of the unique orthonormal frame associated with a  Newman-Penrose (NP) frame adapted to the PNDs, those quantities can be in turn be expressed in terms of the NP curvature scalars, 
and one then has a way of visualizing the properties of the PNDs through their spatial light ray directions and the given observer.

By choosing appropriate observers and frames in an explicit spacetime which is an exact solution of Einstein's equations, one can simplify the description of the PND wedge products and their contractions to relate them more easily to the geometrical properties of that spacetime.
Finally, we show that the 2-form wedge product of PNDs in the case $p=2$ satisfies Maxwell-like equations and illustrate this  with explicit vacuum and nonvacuum spacetime examples.  All of these efforts are aimed at showing how the PNDs concretely affect the geometry of the spacetime when viewed by observers which are tied to that geometry.

\section{PNDs and associated $p$-forms}

The four PNDs of the Weyl tensor of a generic type I spacetime are given by
\beq
\label{pnds_general}
K_{i}=l+\lambda_i^*m +\lambda_i \bar m+|\lambda_i|^2 n\,,\quad i=1\ldots 4\,,
\eeq
where the 4 associated eigenvalues $\lambda_i$ are the distinct roots of the following algebraic equation 
\beq
\label{lambda_eq}
\lambda^4 \psi_4-4\lambda^3\psi_3 +6 \lambda^2\psi_2-4\lambda \psi_1+\psi_0=0\,,
\eeq
having chosen an NP frame $\{l, n, m, \bar m\}$ such that $\psi_4\not =0$~\cite{Stephani:2003tm}.
When there are fewer than 4 distinct eigenvalues, the number of roots and their multiplicities determine the Petrov type of the algebraically special spacetimes.

Every NP frame has a standard associated orthonormal frame $\{e_\alpha\}=\{e_0,e_a\}$
($\alpha=0,1,2,3$, $a=1,2,3$) adapted to an observer 4-velocity $u=e_0$ defined by  
\beq
\label{NPvsorthon}
l=\frac{1}{\sqrt{2}}(e_0+e_1)\,, \qquad
n=\frac{1}{\sqrt{2}}(e_0-e_1)\,, \qquad
m=\frac{1}{\sqrt{2}}(e_2+i e_3) \,,
\eeq
so that the PNDs \eqref{pnds_general} can be re-expressed as
\beq
\label{pnds_general2}
K_{i}=\frac1{\sqrt{2}}\left[(1+|\lambda_i|^2)e_0 +(1-|\lambda_i|^2)e_1+2{\rm Re}(\lambda_i)e_2+2{\rm Im}(\lambda_i)e_3\right]\,.
\eeq
These null vectors can be conveniently rescaled as
\beq
K_{i}=\frac1{\sqrt{2}}(1+|\lambda_i|^2)k_i\,,
\eeq
defining a new set of PNDs by
\beq
\label{kidef}
k_i=e_0+\hat\nu_i\,.
\eeq

We will denote by \eqref{kidef} (with $i=1\ldots p$ and $e_0=u$ the 4-velocity of a family of observers) a maximal set of such rescaled distinct PNDs for a given Weyl curvature tensor, which means that the unit spatial vectors $\hat \nu_i$ are distinct.
One can compute their wedge product for the 3 cases  $p=2,3,4$ of more than one distinct PND
\[
\begin{array}{lll}
&\Omega_{1234}=  k_1\wedge k_2\wedge k_3\wedge k_4&\quad\hbox{(type I)}\,,
\\
&\Omega_{123\phantom{4}}=  k_1\wedge k_2\wedge k_3&\quad\hbox{(type II)}\,,
\\
&\Omega_{12\phantom{34}}=  k_1\wedge k_2  &\quad \hbox{(types D, III)}\,.
\end{array}
\]
Hereafter when referring to the wedge of PNDs we always consider them to be represented in the split form $k_i=u+\hat \nu_i$.
Therefore, if a boosted observer is considered, the corresponding PNDs must always be rescaled, in the sense that if $u$ and $U$ are two observers related by a boost 
\beq
u=\gamma(u,U)(U+\nu(u,U))\,,
\eeq
then the rescaled PNDs $k_i^{\rm resc}={k_i}/{\gamma(u,U)}$ are used to define their wedge $p$-forms which we will denote as $\Omega_{1\ldots p}^{\rm resc}$ (with the label \lq\lq resc" eventually replaced by any label identifying the observer).
These forms can be decomposed into their electric and magnetic parts simply by using the temporal and spatial projections associated with $u$. For example, the spatial projection with respect to $u$ is defined (in component form) by the projection operator $P(u)^\alpha{}_\beta=\delta^\alpha{}_\beta +u^\alpha u_\beta$ (signature $-+++$). 
We will use the term ``$p$-form" for both the covariant and contravariant form of antisymmetric tensors of rank $p$.

Any covariant $p$-form $\sot{S}{p}$ can be expressed in terms of the dual 1-forms $\omega^{\alpha}$ associated with our orthonormal frame $e_\alpha$ by
\beq
\sot{S}{p}=\sot{S}{p}_{[\alpha_1\ldots \alpha_p]}\omega^{\alpha_1}\otimes\ldots\otimes\omega^{\alpha_p}\equiv \frac{1}{p!}\sot{S}{p}_{\alpha_1\ldots \alpha_p}\omega^{\alpha_1}\wedge \ldots \wedge \omega^{\alpha_p}\,.
\eeq
This in turn can be represented in terms of its electric and magnetic parts relative to $u$ by
\beq\label{eq:si}
 \sot{S}{p} = u^\flat \wedge \sot{S}{p}^{({\rm E})}(u) + \sot{S}{p}^{({\rm M})}(u)\,,
\eeq
where the $\flat$ symbol denotes the fully covariant form of a tensor, and
$\sot{S}{p}^{({\rm E})}(u)= - u \leftcontract \sot{S}{p}$ 
and $\sot{S}{p}^{({\rm M})}(u) = P(u) \sot{S}{p} $
or explicitly in component form 
\begin{eqnarray}
\left[\sot{S}{p}^{({\rm E})}(u)\right]_{\alpha_1\ldots\alpha_{p-1}}
&=&-u^{\sigma}\sot{S}{p}_{\sigma\alpha_1\ldots\alpha_{p-1}}
\,,\nonumber\\
\left[\sot{S}{p}^{({\rm M})}(u)\right]_{\alpha_1\ldots\alpha_p}
&=&P(u)^{\beta_1}{}_{\alpha_1}\ldots P(u)^{\beta_p}{}_{\alpha_p}\sot{S}{p}_{\beta_1\ldots\beta_p}\,.
\end{eqnarray}
To simplify notation when applied to the various  $p$-forms $\sot{S}{p}=\Omega_{i_1\ldots i_p}$, we will denote these spatial forms by $\sot{S}{p}^{({\rm E})}(u)=E(u)$ and $\sot{S}{p}^{({\rm M})}(u)=B(u)$, where by spatial with respect to $u$ we mean that any index contraction with $u$ gives zero.  See Ref.~\cite{Jantzen:1992rg} for additional details.
Each $p$-form has an obvious scalar invariant obtained by the contraction of the form with itself
\beq
  \frac1{p!} \sot{S}{p}_{\alpha_1\ldots \alpha_p} \sot{S}{p}^{\alpha_1\ldots \alpha_p}=- E(u)_{\alpha_1\ldots \alpha_{p-1}}E(u)^{\alpha_1\ldots \alpha_{p-1}}+{}^{*u}B(u)_{\alpha_1\ldots \alpha_{3-p}}{}^{*u} B(u)^{\alpha_1\ldots \alpha_{3-p}}\,.
\eeq
Note that any 4-form is automatically purely electric. The remaining PND wedge products of 2 or 3 PNDs turn  out to also be ``electric dominated" in the sense that the self-contraction of $E(u)$ (i.e., its magnitude) is greater than the self-contraction of $B(u)$, so that $u$ can be chosen to make the wedge product purely electric with $B(u)=0$. 
In fact as long as the observer 4-velocity belongs to the span of the PNDs, the corresponding $p$-form will be purely electric, explaining the electric dominance for generic observers.

We will explore below the invariants associated with $\Omega_{i_1\ldots i_p}$, as well as several differential relations of the type $\Omega_{i_1\ldots i_p}^{\alpha\beta\ldots }{}_{;\alpha}=J^{\beta\ldots}$, mimicking electromagnetic currents in a gravitoelectromagnetic analogy.

\subsection{$p=2$}

The simplest case is that of two distinct PNDs
\bea
\Omega_{12}&=& k_1\wedge k_2\nonumber\\
&=& (u+\hat \nu_1)\wedge (u+\hat \nu_2)
\nonumber\\
&=&  u\wedge(\hat \nu_2-\hat \nu_1)+\hat \nu_1\wedge  \hat \nu_2 \,,
\eea
(with $\hat \nu_1\not =\hat \nu_2$) 
identifying two spatial vectors with respect to $u$, namely its electric part
\beq
E(u)=\hat \nu_2-\hat \nu_1\,,\qquad 
\eeq
and its magnetic part,  better represented by its spatial dual
\beq
B(u)={}^{*u}[\hat \nu_1\wedge  \hat \nu_2]=\hat \nu_1\times_u \hat \nu_2\,,
\eeq
where the latter operation is taken by using the unit volume spatial 3-form $\eta(u)^{abc}=u_\alpha\eta^{\alpha abc}$ (also used to define the spatial vector product $\times_u$ with respect to $u$), 
\beq
{}^{*u}[\hat \nu_1\wedge  \hat \nu_2]^{a}=\eta(u)_{abc}\hat \nu_1^b  \hat \nu_2^c\equiv [\hat \nu_1\times_u \hat \nu_2]^a\,.
\eeq
Introducing then the nonzero angle $\theta$ between the direction vectors (since $\hat\nu_1\neq \hat\nu_2$)
\beq
\hat \nu_1\cdot \hat \nu_2=\cos \theta\,,
\eeq
one finds
\beq
 |E(u)|= 2|\sin(\theta/2)|\,,\quad |B(u)|= |\sin\theta|=2|\sin(\theta/2)\cos(\theta/2)|\,,
\eeq
so it is obvious that $|E(u)|>|B(u)|$.
Evaluating the two quadratic invariants in terms of this decomposition one immediately finds the condition
\beq
\frac12[{}^*\Omega_{12}]_{\alpha\beta}[\Omega_{12}]^{\alpha\beta}
=E(u)\cdot B(u)=0\,,
\eeq
implying that the electric and magnetic parts cannot be parallel,
while the other invariant turns out to be
\beq
-\frac12[\Omega_{12}]^{\alpha\beta}[\Omega_{12}]_{\alpha\beta}
=E^2(u)-B^2(u)
=4 \sin^4 (\theta/2)> 0\,,
\eeq
(recall $\theta\neq0$) implying that $\Omega_{12}$ is always electrically dominated.
This is obvious because the 2-plane containing any two independent null directions has to be timelike, so the 2-form self-contraction must be negative regardless of signature (as the wedge product of a pair of spacelike and timelike vectors).
The other extreme for this invariant occurs at $\theta=\pi$, i.e., $\hat\nu_2=-\hat\nu_1$, hence
$E(u)=2\hat \nu_1$ and $B(u)=0$, so that
\beq
\label{Omega_12_gen}
\Omega_{12}=-2 u\wedge \hat \nu_1  \,, 
\eeq
and 
\beq
\label{Omega12quadcv}
\frac12[\Omega_{12}]^{\alpha\beta}[\Omega_{12}]_{\alpha\beta}=E(u)^2=-4\,.
\eeq

The latter case $\hat\nu_2=-\hat\nu_1$ corresponds to a special choice of the observer $u$ (and corresponding adapted frame), which will be referred to as the ``center of velocity" (cv) frame of $k_1$ and $k_2$.
In fact, the average $(k_1+k_2)/2 = u+(\hat\nu_1+\hat\nu_2)/2$ is a timelike vector, since
\beq
(k_1+k_2) \cdot (k_1+k_2) = 2 k_1 \cdot k_2 = -2( 1-\cos\theta)=-4\sin^2(\theta/2) <0\,,
\eeq 
while $k_1-k_2=\hat\nu_1-\hat\nu_2$ is obviously spacelike, with the same normalization $2|\sin(\theta/2)|$ as $k_1+k_2$, so define 
the spatial unit vector directions  $\pm\hat\nu^{\rm cv}$ with respect to $U_{12}$ by
\beq
\hat\nu^{\rm cv}
= \frac{(\hat\nu_1-\hat\nu_2)}{ 2|\sin(\theta/2)| }\,.
\eeq
Hence 
\bea
U_{12}&=&\frac{1}{\sqrt{|2 k_1\cdot k_2|}}(k_1+k_2) 
=\frac{1}{|\sin(\theta/2)|}\left[u+\frac12 (\hat\nu_1+\hat\nu_2)\right]
\eea
is the 4-velocity of the center of velocity observer, for which
\beq
k_1=\left|\sin \left(\theta/2\right)\right| (U_{12} +\hat\nu^{\rm cv}) \,,\qquad 
k_2=\left|\sin \left(\theta/2\right)\right| (U_{12}- \hat\nu^{\rm cv})\,,
\eeq
and
\beq
 \Omega_{12}= k_1 \wedge k_2 = -\frac12 (k_1+k_2)\wedge (k_1-k_2)
=-
2\sin^2(\theta/2)\, U_{12} \wedge \hat\nu^{\rm cv}\,.
\eeq
The new rescaled PNDs are then given by
\beq
k_1^{\rm cv}=\frac{k_1}{\left|\sin \left(\theta/2\right)\right|}=U_{12}+\hat\nu^{\rm cv}\,,\qquad
k_2^{\rm cv}=\frac{k_2}{\left|\sin \left(\theta/2\right)\right|}=U_{12}-\hat\nu^{\rm cv}\,,
\eeq
so that $\Omega_{12}^{\rm cv}=k_1^{\rm cv}\wedge k_2^{\rm cv}$ is purely electric with respect to $U_{12}$. 
 
Any observer 4-velocity $u$ in the plane of the two future pointing PNDs will measure equal but opposite spatial projections in its local rest space, but the average of the two PNDs fixes the timelike projection along $u$ of these two vectors to be equal. When an observer does not see the PND spatial directions as equal and opposite, the average of the two PNDs picks out such a center of velocity observer by boosting appropriately along the average spatial direction.

\subsection{$p=3$}

Introducing the compact notation
\beq
\label{notation}
\hat\nu_{ab\ldots}=\hat\nu_a\wedge \hat\nu_b\wedge \ldots\,,
\eeq
for the case of 3 distinct PNDs we have the similar decomposition
\beq
\Omega_{123}=u\wedge (\hat\nu_{12} +\hat\nu_{23}+\hat\nu_{31})
+\hat\nu_{123}
= u\wedge E(u)+ B(u)
\,.
\eeq 
Note that $B(u)$ vanishes when the three spatial velocity vectors are coplanar, i.e., orthogonal to a spatial vector $n$ in the local rest space of $u$ (namely, spatial with respect to $u$). Any boost of $u$ along a timelike direction in this plane leaves them coplanar, so there is  freedom in the choice of $u$ for which $\Omega_{123}$ is purely electric. 

The spatial dual of $E(u)$ is the vector
\beq
{}^{*u}E(u)=\hat\nu_1\times_u \hat\nu_2+\hat\nu_2\times_u \hat\nu_3+\hat\nu_3\times_u \hat\nu_1\,,
\eeq
while that of $B(u)$ is the scalar
\beq
{}^{*u}B(u)=(\hat\nu_1\times_u \hat\nu_2)\cdot  \hat\nu_3 = {}^{*u}E(u) \cdot \hat\nu_3\,,
\eeq
with equivalent permuted expressions.
The quadratic invariant $[\Omega_{123}]^2$ is given by
\beq
\label{Omega123quad}
\frac{1}{2!3!}[\Omega_{123}
]_{\alpha\beta\gamma}
[\Omega_{123}
]^{\alpha\beta\gamma}
=-
(1-\hat \nu_3\cdot \hat \nu_1)
(1-\hat \nu_2\cdot \hat \nu_1)
(1-\hat \nu_2\cdot \hat \nu_3)\,.
\eeq

In this case we have three different center of velocity frames: 
the center of velocity frame of $k_1$ and $k_2$, $U_{12}={(k_1+k_2)}/{\sqrt{|2 k_1\cdot k_2|}}$,
and analogously $U_{13}$ and $U_{23}$.
For example, in the center of velocity frame for $k_1$ and $k_2$, as  seen above,  we have $\hat\nu_2=-\hat\nu_1$
(or, more properly $\hat\nu_2\equiv \hat\nu(k_2,U_{12})=-\hat\nu(k_1,U_{12})\equiv -\hat\nu_1$ according to the notation introduced in Ref.~\cite{Jantzen:1992rg}), so that 
\beq
{}^{*U_{12}}E(U_{12})=2 \hat\nu_2 \times_u \hat\nu_3 \,,\qquad
{}^{*U_{12}}B(U_{12})=0\,,
\eeq
i.e.,  $B(U_{12})=0$ and the new $\Omega_{123}^{\rm cv}$ is purely electric.
The quadratic invariant \eqref{Omega123quad} then reduces to
\beq
\frac{1}{12}[\Omega_{123}^{\rm cv}]_{\alpha\beta\gamma}[\Omega_{123}^{\rm cv}]^{\alpha\beta\gamma}=-2[1-(\hat \nu_3\cdot \hat \nu_1)^2]\,.
\eeq

\subsection{$p=4$}

Finally, in the  case of 4 distinct PNDs where ${}^{*u}B(u)=\hat\nu_1\wedge \hat\nu_2 \wedge \hat\nu_3\wedge \hat \nu_4=0$, since they belong to a 3-dimensional space, we have $B(u)=0$ and
\beq
\Omega_{1234}=u\wedge E(u)\,,
\eeq 
with
\beq
E(u)= \hat\nu_{124}   +\hat\nu_{234}+\hat\nu_{314}-\hat\nu_{123}\,.
\eeq 

Moreover, the spatial dual of the electric part is a scalar and it is expressed as follows
\beq
{}^{*u}E(u)=(\hat\nu_1\times_u \hat\nu_2)\cdot \hat \nu_4 +(\hat\nu_2\times_u \hat\nu_3)\cdot \hat \nu_4
+(\hat\nu_3\times_u \hat\nu_1)\cdot  \hat \nu_4-(\hat\nu_1\times_u\hat\nu_2)\cdot  \hat \nu_3\,.
\eeq
The quadratic invariant $[\Omega_{1234}]^2$ is then given by
\beq
\label{Omega1234quad}
\frac{1}{2!3!4!}[\Omega_{1234}
]_{\alpha\beta\gamma\delta}[\Omega_{1234}
]^{\alpha\beta\gamma\delta}
=-\frac{1}{12}[{}^{*u}E(u)]^2\,.
\eeq
The previous expressions simplify when evaluated in a center of velocity frame.
For example, with $u=U_{12}$ we find
\beq
{}^{*U_{12}}E(U_{12})=2(\hat\nu_3\times_u \hat\nu_1)\cdot  \hat \nu_4\,.
\eeq

\section{Explicit examples}

In this section we discuss the wedge product $p$-form $\Omega_{\alpha_1\ldots \alpha_p}$ ($p=2,3,4$) of the PNDs in a number of explicit spacetimes of various Petrov types,
starting with $p=2$.
Because of the fundamental reliance of this the Petrov classification quantities on the Newman-Penrose formalism~\cite{Chandrasekhar:1985kt},  we now  switch the metric signature from $-+++$ to $+---$.
The NP frame $\{l, n, m, \bar m\}$ has an associated orthonormal frame $\{e_\alpha\}=\{e_0,e_a\}$ adapted to $u=e_0$
as in Eq.~\eqref{NPvsorthon}, with dual frame $\{\omega^\alpha\}$, and the shorthand notation 
$\omega^{ab\ldots}=\omega^a\wedge \omega^b\wedge\ldots$ will be used.
We decompose the PNDs along frames adapted to various families of observers: a canonical frame, a center of velocity frame, and frames adapted to observers who play a special role in the spacetimes under consideration.

\subsection{Type D spacetimes}

In a canonical NP frame $\{e_\alpha^{\rm can}\}$ such that the Weyl scalars satisfy the conditions 
\beq
\psi_0=\psi_1=\psi_3=\psi_4=0\,,\qquad
\psi_2\not=0\,,
\eeq
the PNDs are simply given by $K_1=l$ and $K_2=n$, so that 
\beq
k_1^{\rm can}=e_0^{\rm can}+e_1^{\rm can}\,,\qquad
k_2^{\rm can}=e_0^{\rm can}-e_1^{\rm can}\,,
\eeq
and the canonical frame is also a center of velocity frame, i.e., $k_i^{\rm can}=k_i^{\rm cv}$.
Taking the wedge product of the latter then yields
\beq
\Omega_{12}^{\rm cv}=-2 e_0\wedge e_1  \,,
\eeq
with spacetime dual
\beq
{}^*\Omega_{12}^{\rm cv}=-2e_2 \wedge e_3 \,.
\eeq
The associated quadratic invariant is then
\beq
\frac12[\Omega_{12}^{\rm cv}]^{\alpha\beta}[\Omega_{12}^{\rm cv}]_{\alpha\beta}=-4\,,
\eeq
according to Eq. \eqref{Omega12quadcv}.

\subsubsection{Kerr spacetime}

Consider the Kerr spacetime, with line element written in standard Boyer-Lindquist coordinates $(t,r,\theta,\phi)$
\beq
ds^2=dt^2 -
\frac{2Mr}{\Sigma} (a \sin^2\theta d\phi - dt)^2 - (r^2 + a^2) \sin^2 \theta d\phi^2
-\frac{\Sigma}{\Delta}dr^2 - \Sigma d\theta^2\,,
\eeq
where
\beq
\Sigma = r^2 + a^2 \cos^2\theta\,, \qquad  \Delta= r^2 + a^2 - 2Mr\,.
\eeq
A canonical NP frame is given by the Kinnersley tetrad \cite{Kinnersley:1969zza}
\bea
l&=&\frac{1}{\Delta}\left[(r^2+a^2)\partial_t+\Delta\partial_r+a\partial_\phi\right]
\,,\nonumber\\
n&=&\frac{1}{2\Sigma}\left[(r^2+a^2)\partial_t-\Delta\partial_r+a\partial_\phi\right]
\,,\nonumber\\
m&=&\frac{1}{\sqrt{2}(r+ia\cos\theta)}\left[ia\sin\theta\partial_t+\partial_\theta+\frac{i}{\sin\theta}\partial_\phi\right]
\,,
\eea
with nonvanishing Weyl scalar 
\beq
\psi_2=-\frac{M}{(r-ia\cos\theta)^3}\,.
\eeq
The type III null rotation (leaving $\psi_2$ unchanged) $l\to {\mathcal A}^{-1}l\,,\quad n\to {\mathcal A}n \,,\quad  m \to e^{i\theta}m\,,\qquad \bar m\to e^{-i\theta}\bar m$, with ${\mathcal A}=\sqrt{2\Sigma/\Delta}$ and $\theta=0$, allows for expressing the PNDs in the form 
\beq
\label{k12car}
k_{1}^{\rm cv}=u^{\rm car}+ e_{\hat r}\,,\qquad 
k_{2}^{\rm cv}=u^{\rm car}- e_{\hat r}\,,
\eeq
in terms of the Carter observers with 4-velocity
\beq
e_0=u^{\rm car}=\frac{r^2+a^2}{\sqrt{\Delta\Sigma}}\left(\partial_t+\frac{a}{r^2+a^2}\partial_\phi\right)
\,,
\eeq
and associated spatial triad 
\beq
e_1
=e_{\hat r}=\sqrt{\Delta/\Sigma}\,\partial_r\,,\
e_2
=e_{\hat \theta}=\sqrt{1/\Sigma}\,\partial_\theta\,,\
e_3
=\bar u^{\rm car}=\frac{a\sin\theta}{\sqrt{\Sigma}}\left(\partial_t+\frac{1}{a\sin^2\theta}\partial_\phi\right)\,.
\eeq
Carter observers thus define a center of velocity frame for $k_1$ and $k_2$, with $\hat \nu_1=e_{\hat r}$ and $\hat \nu_2=-\hat \nu_1=-e_{\hat r}$, so that their wedge product is  
\beq
\Omega_{12}^{\rm cv}=-2 u^{\rm car}\wedge e_{\hat r}  \,,
\eeq
with spacetime dual
\beq
{}^*\Omega_{12}^{\rm cv}=-2e_{\hat \theta} \wedge \bar u^{\rm car} \,.
\eeq
The nonvanishing coordinate components of $\Omega_{12}^{\rm cv}$ and ${}^*\Omega_{12}^{\rm cv}$ are $[\Omega_{12}^{\rm cv}]_{tr}=2$ and $[\Omega_{12}^{\rm cv}]_{r\phi}=2a\sin^2\theta$, and $[{}^*\Omega_{12}^{\rm cv}]_{t\theta}=-2a\sin\theta$ and $[{}^*\Omega_{12}^{\rm cv}]_{\theta\phi}=-2(r^2+a^2)\sin\theta$, respectively.
The associated quadratic invariant has then value $-4$.

Note that the center of velocity frame is not unique. For example, in this case any other observer $U$ obtained by boosting the Carter observer $u^{\rm car}$  in the radial direction (denoting as $E_{\hat r}$ the boosted vector of $e_{\hat r}$) still defines a center of velocity frame. In fact, the relative decomposition of the two families of observers reads
\bea
 u^{\rm car}&=&\gamma(u^{\rm car},U)[U+||\nu(u^{\rm car},U)|| E_{\hat r}]
\,,\nonumber\\ 
e_{\hat r}&=&\gamma(u^{\rm car},U)[||\nu(u^{\rm car},U)|| U+E_{\hat r} ]\,,
\eea
with $\gamma(u^{\rm car},U)=(1-||\nu(u^{\rm car},U)||^2)^{-1/2}$, so that
\beq
u^{\rm car}\wedge e_{\hat r}= U \wedge E_{\hat r}\,, 
\eeq
and 
\bea
k_{1,2}^{\rm cv}&=&u^{\rm car}\pm e_{\hat r}
= {\mathcal A}_\pm \, (U\pm E_{\hat r} )\,,
\eea
with $1\to +$ and $2\to -$, and
\beq
{\mathcal A}_+=\sqrt{\frac{1+ ||\nu(u^{\rm car},U)||}{1- ||\nu(u^{\rm car},U)||}}=\frac{1}{{\mathcal A}_-}\,.
\eeq
However, the value of the associated quadratic invariant does not change.
In fact, the rescaled PNDs are $\tilde k_{1,2}^{\rm cv}= U\pm E_{\hat r}$, and their wedge product turns out to be $\tilde k_1^{\rm cv}\wedge \tilde k_2^{\rm cv}=({\mathcal A}_+{\mathcal A}_-)^{-1}[k_1^{\rm cv}\wedge k_2^{\rm cv}]=k_1^{\rm cv}\wedge k_2^{\rm cv}$.

It is also interesting to consider the contraction of $\Omega_{12}^{\rm cv}$ and ${}^*\Omega_{12}^{\rm cv}$ with the Riemann (Weyl) tensor and its dual, leading to
\bea
\label{rel_C_Omega}
C_{\alpha\beta}{}^{\gamma\delta}[\Omega_{12}^{\rm cv}]_{\gamma\delta}&=&
-{}^*C_{\alpha\beta}{}^{\gamma\delta}[{}^*\Omega_{12}^{\rm cv}]_{\gamma\delta}
=[c_1\,\Omega_{12}^{\rm cv}+c_2{}^*\Omega_{12}^{\rm cv}]_{\alpha\beta}
\,,\nonumber\\
{}^*C_{\alpha\beta}{}^{\gamma\delta}[\Omega_{12}^{\rm cv}]_{\gamma\delta}&=&
C_{\alpha\beta}{}^{\gamma\delta}[{}^*\Omega_{12}^{\rm cv}]_{\gamma\delta}
=\left[-c_2\,\Omega_{12}^{\rm cv}+c_1{}^*\Omega_{12}^{\rm cv}\right]_{\alpha\beta}
\,,
\eea
where 
\bea
c_1&=&-\frac18C_{\alpha\beta\gamma\delta}[\Omega_{12}^{\rm cv}]^{\alpha\beta}[\Omega_{12}^{\rm cv}]^{\gamma\delta}
=-\frac{4Mr}{\Sigma^3}(r^2 -3 a^2 \cos^2\theta)
\,,\nonumber\\
c_2&=&\frac18{}^*C_{\alpha\beta\gamma\delta}[\Omega_{12}^{\rm cv}]^{\alpha\beta}[\Omega_{12}^{\rm cv}]^{\gamma\delta}
=-\frac{4aM\cos\theta}{\Sigma^3}(3r^2 -a^2 \cos^2\theta)
\,.
\eea
Moreover, defining the self-dual combination 
\beq
\hook \Omega_{12}^{\rm cv}=\Omega_{12}^{\rm cv}+i{}^*\Omega_{12}^{\rm cv}\,,
\eeq
and the analogous combination  $\hook C_{\alpha\beta\gamma\delta} = C_{\alpha\beta\gamma\delta}+i{}^*C_{\alpha\beta\gamma\delta}$ for the Weyl tensor, the real relations \eqref{rel_C_Omega} (as well as the coefficients $c_1$ and $c_2$) collapse into a single complex one
\beq
\hook C_{\alpha\beta\gamma\delta}[\hook \Omega_{12}^{\rm cv}]^{\gamma\delta}
=-8\psi_2[\hook \Omega_{12}^{\rm cv}]_{\alpha\beta}\,.
\eeq

Finally, the $2$-form $\Omega_{12}^{\rm cv}$ satisfies the following electromagnetic-like equations
\beq
[\Omega_{12}^{\rm cv}]^{\alpha\beta}{}_{;\beta}=J_e^\alpha\,,\qquad
[{}^*\Omega_{12}^{\rm cv}]^{\alpha\beta}{}_{;\beta}=J_m^\alpha\,,
\eeq
with
\beq
\frac14J_e=-\frac{r}{\Sigma}\partial_t
\equiv\tilde \rho_e\partial_t\,,\qquad
\frac14J_m=\frac{a\cos\theta}{\Sigma}\partial_t
\equiv\tilde \rho_m\partial_t\,,
\eeq
so that 
\beq
\tilde \rho_e^2+\tilde \rho_m^2=1\,.
\eeq
Passing to $[\hook \Omega_{12}^{\rm cv}]$ the above relations become
\beq
[\hook\Omega_{12}^{\rm cv}]^{\alpha\beta}{}_{;\beta}=[J_e+iJ_m]^\alpha
=-\frac{4}{r+ia\cos\theta}\delta^\alpha_0\,,
\eeq
implying that the rescaled quantity $(r+ia\cos\theta)^{-2}\hook\Omega_{12}^{\rm cv}$ is divergence-free, i.e., 
in coordinate components
\bea
[(r+ia\cos\theta)^{-2}\hook\Omega_{12}^{\rm cv}]^{\alpha\beta}{}_{;\beta}&=&
[(r+ia\cos\theta)^{-2}]_{,\beta}[\hook\Omega_{12}^{\rm cv}]^{\alpha\beta}
+(r+ia\cos\theta)^{-2}[\hook\Omega_{12}^{\rm cv}]^{\alpha\beta}_{;\beta}\nonumber\\
&=&4(r+ia\cos\theta)^{-3}\delta^\alpha_0-4(r+ia\cos\theta)^{-3}\delta^\alpha_0
=0\,.
\eea

In the Kerr spacetime there are other observer families naturally associated with its special geometry: the threading (or static) observers following the integral curves of the stationary Killing vector field $\partial_t$, and the zero angular momentum observers (ZAMOs), whose world lines are orthogonal to the time coordinate hypersurfaces.
The static observers exist only in the spacetime region outside the black hole ergosphere where $g_{tt}<0$, and have 4-velocity
\beq
u^{\rm thd}
=\left(1-\frac{2Mr}{\Sigma}\right)^{-1/2}\,\partial_t\,,
\eeq
with adapted frame 
\beq
e_1^{\rm thd} = e_{\hat r}\,,\quad
e_2^{\rm thd} = e_{\hat \theta}\,,\quad
e_3^{\rm thd}=\frac{\sqrt{\Delta -a^2 \sin^2\theta}}{\sin \theta \sqrt{\Delta \Sigma}} \left(\partial_\phi-\frac{2Mar\sin^2\theta}{\Delta -a^2 \sin^2\theta} \partial_t  \right)\,. 
\eeq
The ZAMOs instead exist everywhere outside of the outer horizon $r>r_+=M+\sqrt{M^2-a^2}$, and have 4-velocity 
\beq
u^{\rm zamo}
= \sqrt{\frac{A}{\Delta\Sigma}}\,\left(\partial_t+\frac{2aMr}{A}\partial_\phi\right)\,,
\eeq
with $A=(r^2+a^2)^2-a^2\Delta\sin^2\theta$, and adapted frame
\beq
e_1^{\rm zamo} = e_{\hat r}\,,\quad
e_2^{\rm zamo} = e_{\hat \theta}\,,\quad
e_3^{\rm zamo}=\frac{\sqrt{\Sigma}}{\sin\theta \sqrt{A}} \,\partial_\phi
\equiv e_{\hat \phi}\,. 
\eeq

Carter observers, static observers and ZAMOs all share the same $r$-$\theta$ 2-plane of their local rest spaces.
They differ only by relative azimuthal motion, so that their adapted frames are all related by relative boosts in the $t$-$\phi$ plane of the tangent space.
The relative decomposition of the Carter observers with respect to the static observers is then 
\beq
u^{\rm car}=\gamma(u^{\rm car},u^{\rm thd})\left[u^{\rm thd}+\nu(u^{\rm car},u^{\rm thd})\right]\,,
\eeq
with
\beq
\gamma(u^{\rm car},u^{\rm thd})=\left(\frac{\Delta}{\Delta -a^2 \sin^2\theta}\right)^{1/2}\,,\qquad
\nu(u^{\rm car},u^{\rm thd})= \frac{a\sin\theta}{\sqrt{\Delta}}e_3^{\rm thd}\,,
\eeq
whereas with respect to ZAMOs is 
\beq
u^{\rm car}=\gamma(u^{\rm car},u^{\rm zamo})\left[u^{\rm zamo}+\nu(u^{\rm car},u^{\rm zamo})\right]\,,
\eeq
with
\beq
\gamma(u^{\rm car},u^{\rm zamo})=\frac{r^2+a^2}{\sqrt{A}}\,,\qquad
\nu(u^{\rm car},u^{\rm zamo})= -\frac{a\sin\theta\sqrt{\Delta}}{r^2+a^2}e_3^{\rm zamo}\,.
\eeq
One can then decompose the PNDs with respect to these new families of observers simply by rescaling those corresponding to the Carter observers, Eq. \eqref{k12car}, by the corresponding $\gamma$ factors.
As a result, the wedge product of the new PNDs is rescaled by a factor of $\gamma^2$, and the associated quadratic invariant by a factor of $\gamma^4$, so that
\bea
\frac12[\Omega_{12}^{\rm thd}]^{\alpha\beta}[\Omega_{12}^{\rm thd}]_{\alpha\beta}&=&-\frac4{\gamma(u^{\rm car},u^{\rm thd})^4}
\,,\nonumber\\
\frac12[\Omega_{12}^{\rm zamo}]^{\alpha\beta}[\Omega_{12}^{\rm zamo}]_{\alpha\beta}&=&-\frac4{\gamma(u^{\rm car},u^{\rm zamo})^4}
\,.
\eea

\subsubsection{Kasner spacetime}

Consider the vacuum Kasner~\cite{Landau:1975pou} spacetime 
\begin{equation}
ds^2=dt^2-t^{2p_1}d x^2-t^{2p_2}dy^2-t^{2p_3}dz^2\,,  \label{LLLL}
\end{equation}
where the so-called Kasner indices $p_i$ satisfy $p_1+p_2+p_3=p_1^2+p_2^2+p_3^2=1$
and assume values in the closed interval $[-\frac13,1]$.
The generic Petrov type is I.
In fact, introducing the following NP frame
\begin{eqnarray}  
\label{tetrade}
l = \frac{1}{\sqrt{2}}[\partial_t+t^{-p_1}\partial_x]\,, \quad
  n=\frac{1}{\sqrt{2}}[\partial_t-t^{-p_1}\partial_x] \,,  \quad
m =\frac 1{\sqrt{2}}[t^{-p_2}\partial_y+it^{-p_3}\partial_z] \,,  
\end{eqnarray}
the nonzero Weyl scalars are
\begin{eqnarray}
\label{weylscal}
&&\psi _0=\psi_4=\frac{p_1(p_2-p_3)}{2t^2}, \quad \psi _2=-\frac{p_2p_3}{2t^2}\,.
\end{eqnarray}
The special case $p_1=1,p_2=p_3=0$ (and permutations) corresponds to flat spacetime, while the general type I Kasner case was analyzed in \cite{Bini:2023}.

Here we consider the special case $p_1=-1/3$, $p_2=p_3=2/3$ (and permutations) which corresponds to the type D Kasner model, with a spindle-like cosmological singularity~\cite{Stephani:2003tm,DIN}. The null tetrad above is a principal one in this case, with $l$ and $n$ aligned along the PNDs of the Weyl tensor.
The orthonormal frame naturally associated with \eqref{tetrade}
\beq\label{orto}
e_0=\partial_t\,,\quad
e_1=t^{-p_1}\partial_x\,,\quad
e_2=t^{-p_2}\partial_y\,,\quad
e_3=t^{-p_3}\partial_z\,,
\eeq
is adapted to the static observers with 4-velocity $u=e_0$ whose spatial axes are aligned with the Killing vectors $\partial_x, \partial_y, \partial_z$, and therefore directly observe the homogeneity of the spacetime.
The static observers define the center of velocity frame of $k_1$ and $k_2$, which can then be written as
\beq
k_1^{\rm cv}=u+e_1\,,\qquad
k_2^{\rm cv}=u-e_1\,,
\eeq
with associated wedge product $2$-form $\Omega_{12}^{\rm cv}$ and its spacetime dual ${}^*\Omega_{12}^{\rm cv}$ given by
\beq
\Omega_{12}^{\rm cv}=-2u\wedge e_1  \,,\qquad
{}^*\Omega_{12}^{\rm cv}=-2e_2\wedge e_3 \,.
\eeq
The quadratic invariant has then value $-4$.
The contraction of $\Omega_{12}^{\rm cv}$ and ${}^*\Omega_{12}^{\rm cv}$ with the Weyl tensor (and its dual) are 
\bea
C_{\alpha\beta}{}^{\gamma\delta}[\Omega_{12}^{\rm cv}]_{\gamma\delta}&=&
-{}^*C_{\alpha\beta}{}^{\gamma\delta}[{}^*\Omega_{12}^{\rm cv}]_{\gamma\delta}
=c_1[\Omega_{12}^{\rm cv}]_{\alpha\beta}
\,,\nonumber\\
{}^*C_{\alpha\beta}{}^{\gamma\delta}[\Omega_{12}^{\rm cv}]_{\gamma\delta}&=&
C_{\alpha\beta}{}^{\gamma\delta}[{}^*\Omega_{12}^{\rm cv}]_{\gamma\delta}
=c_1[{}^*\Omega_{12}^{\rm cv}]_{\alpha\beta}\,,
\eea
where
\beq
c_1=-\frac18C_{\alpha\beta\gamma\delta}[\Omega_{12}^{\rm cv}]^{\alpha\beta}[\Omega_{12}^{\rm cv}]^{\gamma\delta}
=\frac{2p_1(1-p_1)}{t^2}
=-\frac{2p_2p_3}{t^2}
\,.
\eeq

Finally, the $2$-form $\Omega_{12}^{\rm cv}$ satisfies the following electromagnetic-like equations
\beq
[\Omega_{12}^{\rm cv}]^{\alpha\beta}{}_{;\beta}=J^\alpha\,,\qquad
[{}^*\Omega_{12}^{\rm cv}]^{\alpha\beta}{}_{;\beta}=0\,,
\eeq
with
\beq
J=\frac{2(1-p_1)}{t}e_1\,,
\eeq
so that the rescaled quantity $t^{-(1-p_1)}\Omega_{12}^{\rm cv}$ is divergence-free.

\subsection{Type II spacetimes}

Type II spacetimes are rather unfamiliar, because they are not associated with well-known or astrophysically relevant  gravitational fields. 
We consider below two nonvacuum solutions belonging to the Robinson-Trautman class admitting a null vector field that is vorticity-free and shear-free, but with nonzero expansion (see, e.g., Ref. \cite{Podolsky:2016sff} for a complete algebraic classification of such spacetimes).
Their line elements are of the following form
\beq
ds^2=2H du^2+2du dr -A(dx^2+dy^2)\,,
\eeq
depending on two positive metric functions $H(u,r,x,y)$ and $A(u,r,x,y)$ of the coordinates.

One can always choose a canonical NP frame (with associated orthonormal frame $\{e_\alpha^{\rm can}\}$ adapted to $u^{\rm can}=e_0^{\rm can}$), corresponding to $\psi_0=0=\psi_1=\psi_3$ and $\psi_4=-2$ \cite{Stephani:2003tm}, so that Eq.~\eqref{lambda_eq} becomes
\beq
\label{lambda_eq_II}
-2\lambda^2(\lambda^2-3\psi_2)=0\,,
\eeq
with solutions $(\psi_2\neq0)$
\beq
\label{lambdasIIcanframe}
\lambda_1=0=\lambda_2\,, \quad 
\lambda_3=\sqrt{3\psi_2}\,, \quad
\lambda_4=-\lambda_3
\,.
\eeq 
Therefore, $K_1=l=K_2$ is a repeated PND with multiplicity 2, while $K_3,K_4$ are given by Eq.~\eqref{pnds_general}.
The rescaled PNDs $k_i^{\rm can}=u^{\rm can}+\hat\nu_i^{\rm can}$ then have the unit spatial vectors
\bea
\label{hatnuatypeII}
\hat\nu_1^{\rm can}&=&e_1^{\rm can}
\,,\nonumber\\
\hat\nu_2^{\rm can}&=&\sin\alpha\cos\beta e_1^{\rm can}+\sin\alpha\sin\beta e_2^{\rm can}+\cos\alpha e_3^{\rm can}
=\hat\nu_2^{\rm can}(\alpha,\beta)
\,,\nonumber\\
\hat\nu_3^{\rm can}&=&
\hat\nu_2^{\rm can}(\pi-\alpha,-\beta)
\,,
\eea
with
\beq
\beta={\rm arctan}\left(\frac{2{\rm Re}(\lambda_3)}{1-|\lambda_3|^2}\right)\,,\qquad
\alpha={\rm arccos}\left(\frac{2{\rm Im}(\lambda_3)}{1+|\lambda_3|^2}\right)\,.
\eeq
The quadratic invariant \eqref{Omega123quad} is then given by
\beq
\label{Omega123quad2}
\frac{1}{2!3!}[\Omega_{123}^{\rm can}]_{\alpha\beta\gamma}[\Omega_{123}^{\rm can}]^{\alpha\beta\gamma}=\frac{32|\lambda_3|^6}{(1+|\lambda_3|^2)^4}\,.
\eeq

\subsubsection{Robinson-Trautman spacetime with a scalar field source}

A Petrov type II solution belonging to the Robinson-Trautman class with a minimally coupled massless scalar field $\phi$ was discussed in Refs.~\cite{Tahamtan:2015sra,Tahamtan:2016fur}.
The line element can be conveniently written in the form 
\beq
ds^2=\frac{k+rU'}{U}du^2+2du dr -\frac{r^2U^2-C^2}{Up^2}(dx^2+dy^2)\,,
\eeq
with 
\beq
\phi(u,r)=\frac1{\sqrt{2}}\ln\left(\frac{rU-C}{rU+C}\right)
\,,\qquad
U(u)=\gamma e^{\omega^2 u^2+\eta u} 
\,,
\eeq
and the metric functions $p=p(x,y)$ and $k=k(x,y)$ which satisfy the equations
\beq
\Delta\ln p=k\,,\qquad
\Delta k = 4 C^2\omega^2 \,,
\eeq
where $\Delta\equiv p(x,y)^2(\partial_{xx}+\partial_{yy})$ is the Laplace operator of the transverse 2-space, and $C$, $\gamma$, $\omega$ and $\eta$ are positive constants.
Such a spacetime is of type D in the special case $k=$ const., i.e., when the transverse 2-space has a constant Gaussian curvature.

Choosing the NP frame 
\beq
l=\partial_r\,, \qquad
n=\partial_u-\frac1{2U}(k+rU')\partial_r\,, \qquad
m=\frac{\sqrt{U}p}{\sqrt{2(r^2U^2-C^2)}}(\partial_x+i\partial_y)\,,
\eeq
leads to the following nonvanishing Weyl scalars
\bea
\psi_2&=&\frac13C^2U\frac{k-rU'}{(r^2U^2-C^2)^2}
\,,\nonumber\\
\psi_3&=&\frac12\frac{rU}{r^2U^2-C^2}\bar\delta k
\,,\nonumber\\
\psi_4&=&\frac12\frac1{pU}\bar\delta (p\bar\delta k)
\,,
\eea
where $\bar\delta=\bar m^\alpha\partial_\alpha$.
In the type D case $\psi_3=0=\psi_4$.

The PNDs are $K_1=l$ (with multiplicity 2) and $K_{2,3}\equiv K_{\pm}=l+|\lambda_\pm|^2n+\bar\lambda_\pm m+\lambda_\pm\bar m$, with 
\beq
\lambda_\pm=2\frac{\psi_3}{\psi_4}\pm\left[4\left(\frac{\psi_3}{\psi_4}\right)^2-6\frac{\psi_2}{\psi_4}\right]^{1/2}\,.
\eeq
Rescaling these directions as
\bea
k_1&=&\sqrt{2}K_1
=e_0+e_1
\,,\nonumber\\
k_{2,3}&\equiv& k_\pm = \frac{\sqrt{2}}{1+|\lambda_\pm|^2}K_\pm\nonumber\\
&=&e_0 +\frac{1-|\lambda_\pm|^2}{1+|\lambda_\pm|^2}e_1+\frac{2{\rm Re}(\lambda_\pm)}{1+|\lambda_\pm|^2}e_2+\frac{2{\rm Im}(\lambda_\pm)}{1+|\lambda_\pm|^2}e_3
\,,
\eea
leads to
\bea
\Omega_{123}&=&C_1\omega^{012}+C_2\omega^{013}+C_3(\omega^{023}+\omega^{123})\nonumber\\
&=&\omega^0\wedge(C_1\omega^{12}+C_2\omega^{13}+C_3\omega^{23})+C_3\omega^{123}\nonumber\\
&=&\omega^{01}\wedge (C_1\omega^2+C_2\omega^3)+C_3(\omega^0+\omega^1)\wedge \omega^{23}\,,
\eea
with
\bea
C_1&=&-\frac{4\left[{\rm Re}(\lambda_-)|\lambda_+|^2-{\rm Re}(\lambda_+)|\lambda_-|^2\right]}{(1+|\lambda_+|^2)^2(1+|\lambda_-|^2)^2}
\,,\nonumber\\
C_2&=&\frac{4\left[{\rm Im}(\lambda_-)|\lambda_+|^2-{\rm Im}(\lambda_+)|\lambda_-|^2\right]}{(1+|\lambda_+|^2)^2(1+|\lambda_-|^2)^2}
\,,\nonumber\\
C_3&=&-\frac{4\left[{\rm Re}(\lambda_+){\rm Im}(\lambda_-)-{\rm Re}(\lambda_-){\rm Im}(\lambda_+)\right]}{(1+|\lambda_+|^2)^2(1+|\lambda_-|^2)^2}\,.
\eea
The quadratic invariant $[\Omega_{123}]^2$ is then given by
\beq
\frac{1}{2!3!}[\Omega_{123}]_{\alpha\beta\gamma}[\Omega_{123}]^{\alpha\beta\gamma}
=\frac{8|\lambda_+|^2|\lambda_-|^2|\lambda_+-\lambda_-|^2}{(1+|\lambda_+|^2)^2(1+|\lambda_-|^2)^2}\,.
\eeq
The contraction of $\Omega_{123}$ with the Weyl tensor is identically zero.

\subsubsection{Bonnor-Davidson solution with a perfect fluid source}

The Bonnor-Davidson solution describes a stationary nonvacuum spacetime filled with a perfect fluid  having nonzero vorticity and obeying the equation of state $\rho+3p=$ const. \cite{Bonnor:1985}.
The line element in $(u,r,x,y)$ coordinates is
\beq
ds^2=2Hdu^2+2dudr-A(dx^2+dy^2)\,,
\eeq
where $H$ and $A$ are (positive) functions of $r$, $x$, $y$ given by
\beq
H=-3x+n[1-(kr+m)\cot kr]\,, \qquad
A=\frac{\sin^2kr}{4kx^3}\,,
\eeq 
with arbitrary parameters  $(m,k,n)$.
The fluid 4-velocity $U=1/(\sqrt{2H})\partial_u$ is tangent to the Killing vector $\partial_u$.
The energy density $\rho$ and pressure $p$ satisfy the relations
\beq
2\pi(\rho+3p)=nk^2\,, \qquad
2\pi(\rho+p)=k^2H\,.
\eeq 
The positivity of the metric functions implies that for fixed values of the parameters the allowed ranges for the coordinates $r$ and $x$ are determined by the conditions $0<x<x_{\rm max}(r)$ and $0<r<r_{\rm max}$, with 
\beq
\label{x_max}
x_{\rm max}=\frac{n}{3}[1-(kr+m)\cot kr]\,.
\eeq 
For example, choosing $n=1=k$ and $m=0$, Eq.~\eqref{x_max} gives $x_{\rm max}=\frac{1}{3}(1-r\cot r)$, so that $x_{\rm max}\to0$ for $r\to0$ and $x_{\rm max}\to\infty$ for $r\to\pi=r_{\rm max}$.

Choosing the NP frame 
\beq
\label{BDframenp}
l=\partial_r\,, \qquad
n=\partial_u-H\partial_r\,, \qquad
m=\frac{1}{\sqrt{2A}}(\partial_x-i\partial_y)\,,
\eeq
leads to the following nonvanishing Weyl scalars
\bea
\psi_2&=&-nk^2\left[\frac13-\frac{1-(kr+m)\cot kr}{\sin^2kr}\right]
\,,\nonumber\\
\psi_3&=&\frac{3\sqrt{2}k^2x^{3/2}\cos kr}{\sin^2kr}
\,,\nonumber\\
\psi_4&=&-\frac{18k^2x^2}{\sin^2kr}
\,.
\eea
The four PNDs are $K_1=l$ (with multiplicity 2) and $K_{2,3}\equiv K_\pm=l+\lambda_\pm^2n+\lambda_\pm(m+\bar m)$, with real eigenvalues
\bea
\lambda_\pm&=&-\frac{\sqrt{2}\cos kr}{3\sqrt{x}}\pm W
\,,\nonumber\\
W&=&\frac{1}{3x}\left[(2x+n)\cos^2kr-3n(kr+m)\cot kr+2n\right]^{1/2}
\,.
\eea
Rescaling these directions as
\bea
k_1&=&\sqrt{2}K_1
=e_0+e_1
\,,\nonumber\\
k_{2,3}&\equiv&k_\pm=\frac{\sqrt{2}}{1+\lambda_\pm^2}K_\pm
=e_0 +\frac{1-\lambda_\pm^2}{1+\lambda_\pm^2}e_1+\frac{2\lambda_\pm}{1+\lambda_\pm^2}e_2\,,
\eea
leads to
\beq
\Omega_{123}=-\frac{4\lambda_+\lambda_-(\lambda_+-\lambda_-)}{(1+\lambda_+^2)(1+\lambda_-^2)}\,\omega^{012}\,.
\eeq
The quadratic invariant $[\Omega_{123}]^2$ is then given by
\beq
\frac{1}{2!3!}[\Omega_{123}]_{\alpha\beta\gamma}[\Omega_{123}]^{\alpha\beta\gamma}
=\frac{8\lambda_+^2\lambda_-^2(\lambda_+-\lambda_-)^2}{(1+\lambda_+^2)^2(1+\lambda_-^2)^2}\,.
\eeq
Moreover, the contraction of $\Omega_{123}$ with the Weyl tensor is identically vanishing.

One can also introduce the center of velocity of $k_1=e_0+e_1$ (i.e., with $\hat \nu_1=e_1$) and $k_2=k_+$
\bea
U_{12}&=&\frac{(1+\lambda_+^2)^{1/2}}{\lambda_+}\left[
e_0+\frac{1}{1+\lambda_+^2}(e_1+\lambda_+ e_2)
\right]\nonumber\\
&=& \gamma_{12}\left[e_0+v_{12} \hat n_{12}\right]\,,
\eea
where $\gamma_{12}=(1-v_{12}^2)^{-1/2}$ and
\beq
\label{hatn12def}
v_{12}=\frac{1}{\sqrt{1+\lambda_+^2}}\,,\qquad
\hat n_{12}= \frac{1}{\sqrt{1+\lambda_+^2}}e_1+\frac{\lambda_+}{\sqrt{1+\lambda_+^2}} e_2\,,
\eeq
i.e., $U_{12}$ is obtained by boosting $e_0$ along the direction $\hat n_{12}$.
[Eq. \eqref{hatn12def} also suggests the parametrization $\lambda_+=\tan \sigma$ leading to simple trigonometric expressions for the above coefficients.]
An adapted spatial triad to $U_{12}$ is
\bea
E_1&=&\hat\nu_1^{\rm cv}=\frac1{(1+\lambda_+^2)^{1/2}}(\lambda_+ e_1-e_2)
=-\hat\nu_2^{\rm cv}\,,\nonumber\\
E_2&=&\gamma_{12}(v_{12}e_0 +\hat n_{12})=\frac1{\lambda_+}(e_0+e_1+\lambda_+ e_2)\,,\nonumber\\
E_3&=&e_3\,.
\eea
We find then
\beq
k_1^{\rm cv}=\gamma_{12}k_1=U_{12}+\hat\nu_1^{\rm cv}\,,\qquad 
k_2^{\rm cv}=\gamma_{12} k_2=U_{12}-\hat\nu_1^{\rm cv}\,,
\eeq
so that 
\beq
\Omega_{123}^{\rm cv}=k_1^{\rm cv}\wedge k_2^{\rm cv}
=-\frac{4\lambda_-(\lambda_+-\lambda_-)}{\lambda_+(1+\lambda_-^2)}\,U_{12}\wedge \hat\nu_1^{\rm cv}\wedge E_2\,.
\eeq
The quadratic invariant $[\Omega_{123}^{\rm cv}]^2$ is then given by
\beq
\frac1{2!3!}[\Omega_{123}^{\rm cv}]_{\alpha\beta\gamma}[\Omega_{123}^{\rm cv}]^{\alpha\beta\gamma}=\frac{8\lambda_-^2(\lambda_+-\lambda_-)^2}{\lambda_+^2(1+\lambda_-^2)^2}\,.
\eeq

Finally, one can equivalently pass to a canonical orthonormal frame $\{e_\alpha^{\rm can}\}$ by successively applying to the NP frame \eqref{BDframenp} a type I null rotation $l\to l\,,\quad m \to m+al \,,\quad \bar m\to \bar m +\bar a l\,, n\to n+\bar a m + a \bar m + a \bar a l$ (to eliminate $\psi_3$) and a type III null rotation $l\to {\mathcal A}^{-1}l\,,\quad n\to {\mathcal A}n \,,\quad  m \to e^{i\theta}m\,,\qquad \bar m\to e^{-i\theta}\bar m$ (to set $\psi_4=-2$) with (real) respective parameters 
\beq
a=-\frac{\sqrt{2}k^2x^{3/2}\cos kr}{\psi_2\sin^2kr}
\,,\qquad
{\mathcal A}=\frac{\sqrt{3\psi_2}\sin^2kr}{9k^2x^2W}
\,,\qquad
\theta=0\,,
\eeq  
leaving $\psi_2$ unchanged.
With respect to the canonical frame the PNDs $k_i^{\rm can}=u^{\rm can}+\hat\nu_i^{\rm can}$ have unit spatial vectors \eqref{hatnuatypeII}, and quadratic invariant \eqref{Omega123quad2} with $\lambda_3=\sqrt{3\psi_2}$.

\subsection{Type III spacetimes}

A canonical NP frame $\{e_\alpha^{\rm can}\}$ is such that the Weyl scalars are all zero ($\psi_0=\psi_1=\psi_2=\psi_4=0$) except $\psi_3=-i$, so that the PNDs are simply given by $K_1=l$ and $K_2=n$, just as the type D case.
Therefore, 
\beq
k_1^{\rm can}=e_0^{\rm can}+e_1^{\rm can}=k_1^{\rm cv}\,,\qquad
k_2^{\rm can}=e_0^{\rm can}-e_1^{\rm can}=k_2^{\rm cv}\,,
\eeq
and
\beq
\Omega_{12}^{\rm cv}=-2 e_0\wedge e_1  \,,\qquad
{}^*\Omega_{12}^{\rm cv}=-2e_2 \wedge e_3 \,.
\eeq
with associated quadratic invariant equal to $-4$, according to Eq. \eqref{Omega12quadcv}.

\subsubsection{Allnutt spacetime}

Consider the Allnutt spacetime \cite{Allnutt:1981} (see also Ref.~\cite{Munoz:2022duf}), whose metric written in coordinates $(t,u,x,y)$ (with null coordinate $t$) is
\bea
ds^2&=&2e^{2t+3x-y} du(dt+3dx+dy)-\left(3e^{t+2(x-y)} +\frac{1}{a}e^{3(t+2x)}\right)du^2\nonumber\\
&&
-2e^{3t+4x}dx^2-\frac{ae^t}{e^{2y}-1}dy^2\,,
\eea
where $a$ is a constant  positive parameter.
This is a nonvacuum solution of the Einstein equations with a perfect fluid source with (irrotational but shearing and expanding) 4-velocity\footnote{Note a typo in the exact solution book \cite{Stephani:2003tm}, where the second term in the expression for the fluid $4$-velocity (2 lines after Eq. (33.45)) is $e^{-3 t-3 x }$ instead of $e^{-3 t-4 x }$ as in Eq. \eqref{Ufluid}.}
\beq
\label{Ufluid}
U^\flat =\left(\frac{e^{-t}}{a}-\frac32 e^{-3 t-4 x }\right)^{-1/2}dt\,,
\eeq
and energy density and pressure given by
\bea
8\pi \mu &=& \frac{27}{4}\frac{e^{-t}}{a}-\frac98 e^{-3t-4x}\,,\nonumber\\
8\pi p &=& -\frac{21}{4}\frac{e^{-t}}{a}-\frac98 e^{-3t-4x}\,,
\eea
such that $e^{2y}>1$ and $e^{2t+4x}>3a/2$.

An orthonormal co-frame $\{\omega^\alpha\}$ reads
\bea
\omega^0&=&\frac{e^{2 t+3 x-y}}{\sqrt{-g_{uu}}}(dt+3dx+dy)
\,,\nonumber\\
\omega^1&=&\sqrt{-g_{uu}} du-\frac{e^{2 t+3 x-y}}{\sqrt{-g_{uu}}} (dt+3dx+dy)  
\,,\nonumber\\
\omega^2&=& \sqrt{-g_{xx}} dx 
\,,\nonumber\\
\omega^3&=&\sqrt{-g_{yy}} dy\,.
\eea
Passing to the corresponding frame vectors $\{e_\alpha\}$ 
\bea
\label{allnuttframe}
e_0&=&\frac{\sqrt{a}e^{3y/2}}{f^{1/4}(3a+f)^{1/2}}\left[
\frac{e^x(3a+f)}{af^{1/2}}\partial_t+\partial_u
\right]
\,,\nonumber\\
e_1&=&-\frac{\sqrt{a}e^{3y/2}}{f^{1/4}(3a+f)^{1/2}}\partial_u
\,,\nonumber\\
e_2&=&\frac{e^{x+3y/2}}{\sqrt{2}f^{3/4}}\left(3\partial_t-\partial_x\right)
\,,\nonumber\\
e_3&=&\frac{\sqrt{e^{2y}-1}e^{x+y/2}}{a^{1/2}f^{1/4}}\left(\partial_t-\partial_y\right)
\,,
\eea
where $f=e^{2 (t+2 x+ y)}$, and introducing the standard NP frame \eqref{NPvsorthon}, the only onvanishing Weyl scalars are $\psi_3$ and $\psi_4$
\beq
\psi_3
= -\frac34 \sqrt{2a} \frac{e^{2 x+3 y}}{ f^{3/2}(3a+f)^{1/2}}\,,\qquad
\psi_4
= -\frac92 a  \frac{e^{2 x+3 y}}{ f^{3/2}(3a+f)} \,.
\eeq

The two PNDs are $K_1=l$ (with multiplicity 3) and $K_2=l+\lambda^2n+\lambda(m+\bar m)$, with real eigenvalue
\beq
\lambda=-4\frac{\psi_3}{\psi_4}=\frac23\left[\frac{2(3a+f)}{a}\right]^{1/2}\,.
\eeq
Rescaling these directions as
\bea
k_1&=&\sqrt{2}K_1
=e_0+e_1
\,,\nonumber\\
k_2&=&\frac{\sqrt{2}}{1+\lambda^2}K_2
=e_0+\frac{1-\lambda^2}{1+\lambda^2}e_1+\frac{2\lambda}{1+\lambda^2}e_2
\,,
\eea
leads to
\beq
\Omega_{12}=k_1\wedge k_1
=\frac{2\lambda}{1+\lambda^2}\left(-\lambda\omega^{01}+\omega^{02}+\omega^{12}\right)\,,
\eeq
with associated quadratic invariant
\beq
\frac12[\Omega_{12}]_{\alpha\beta}[\Omega_{12}]^{\alpha\beta}=-\frac{4\lambda^4}{(1+\lambda^2)^2}\,.
\eeq

One can also introduce the center of velocity of $k_1$ and $k_2$
\beq
U_{12}=\frac{(1+\lambda^2)^{1/2}}{\lambda}\left[
e_0+\frac{1}{1+\lambda^2}(e_1+\lambda e_2)
\right]
\equiv\gamma_{12}(e_0+v_{12}\hat n_{12})\,.
\eeq
In this frame the new rescaled PNDs $k_1^{\rm cv}$ and $k_2^{\rm cv}$ have unit spatial velocities 
\beq
\hat\nu_1^{\rm cv}=\frac1{(1+\lambda^2)^{1/2}}(\lambda e_1-e_2)
=\sin \sigma e_1-\cos \sigma e_2
=-\hat\nu_2^{\rm cv}\,,
\eeq
with $\lambda=\tan \sigma$, and adapted spatial triad
\beq
E_1=\hat\nu_1^{\rm cv}\,,\qquad
E_2=\frac1{\lambda}(e_0+e_1+\lambda e_2)
\equiv\gamma_{12}(v_{12}e_0+\hat n_{12})\,,\qquad
E_3=e_3\,,
\eeq
so that 
\beq
\Omega_{12}^{\rm cv}=-2U_{12}\wedge \hat\nu_1^{\rm cv}\,,\qquad
{}^*\Omega_{12}^{\rm cv}=2E_2\wedge E_3\,,
\eeq
and the quadratic invariant has value $-4$. 
The contraction of $\Omega_{12}^{\rm cv}$ and its dual with the Weyl tensor gives
\bea
C^{\alpha\beta}{}_{\gamma\delta}[\Omega_{12}^{\rm cv}]^{\gamma\delta}&=&
\frac{4e^{2x+3y}}{f^{3/2}(1+\lambda^2)^{1/2}}\left[k_1^{\rm cv}\wedge E_2\right]^{\alpha\beta}
\,,\nonumber\\
C^{\alpha\beta}{}_{\gamma\delta}[{}^*\Omega_{12}^{\rm cv}]^{\gamma\delta}&=&
\frac{4e^{2x+3y}}{f^{3/2}(1+\lambda^2)^{1/2}}\left[k_1^{\rm cv}\wedge E_3\right]^{\alpha\beta}\,,
\eea
with
\beq
C_{\alpha\beta\gamma\delta}[\Omega_{12}^{\rm cv}]^{\alpha\beta}[\Omega_{12}^{\rm cv}]^{\gamma\delta}=0
=C_{\alpha\beta\gamma\delta}[{}^*\Omega_{12}^{\rm cv}]^{\alpha\beta}[{}^*\Omega_{12}^{\rm cv}]^{\gamma\delta}\,.
\eeq
Finally, the $2$-form $\Omega_{12}^{\rm cv}$ satisfies the following Maxwell-like equations
\beq
[\Omega_{12}^{\rm cv}]^{\alpha\beta}{}_{;\beta}=J_e^\alpha\,,\qquad
[{}^*\Omega_{12}^{\rm cv}]^{\alpha\beta}{}_{;\beta}=J_m^\alpha\,,
\eeq
with spacelike currents
\bea
J_e&=&\frac{e^{x+3y/2}}{\sqrt{2}f^{3/4}(1+\lambda^2)^{1/2}}\left[U_{12}+ (7+6\lambda^2)E_1+3(1+\lambda^2)^{1/2}E_2\right]
\,,\nonumber\\
J_m&=&-\frac{6\sqrt{2}e^{x+3y/2}}{f^{3/4}}\left[-\left(\frac{f(1-e^{-2y})}{2a}\right)^{1/2}E_2+E_3\right]\,.
\eea

A canonical NP frame can be obtained from that associated with the orthonormal frame \eqref{allnuttframe} by performing a type I null rotation $l\to l\,,\quad m \to m+bl \,,\quad \bar m\to \bar m +\bar b l\,, n\to n+\bar b m + b \bar m + b \bar b l$ (to eliminate $\psi_4$) followed by a type III null rotation $l\to {\mathcal A}^{-1}l\,,\quad n\to {\mathcal A}n \,,\quad  m \to e^{i\theta}m\,,\qquad \bar m\to e^{-i\theta}\bar m$ (to set $\psi_3=-i$) with (real) parameters 
\beq
b=-\frac1{\lambda}\,,\qquad
{\mathcal A}=-\lambda f^{3/2}e^{-2 x-3 y}\,,\qquad
\theta=\frac{\pi}{2}\,.
\eeq

\subsection{Type I spacetimes}

In a canonical NP frame such that the Weyl scalars satisfy the conditions 
\beq
\psi_0=\psi_4\not=0\,,\qquad
\psi_2\not=0\,,\qquad
\psi_1=0=\psi_3\,,
\eeq
the four PNDs \eqref{pnds_general} are such that \cite{ArMcIn1994} 
\beq
\label{lambda1sol}
\lambda_1=\left[-3\frac{\psi_2}{\psi_0}-\sqrt{9\left(\frac{\psi_2}{\psi_0}\right)^2-1}\right]^{1/2}\,,
\eeq
and
\beq
\label{lambdasIcanframe}
\lambda_2=-\lambda_1\,, \quad
\lambda_3=\frac1{\lambda_1}\,, \quad
\lambda_4=-\frac1{\lambda_1}\,.
\eeq 
The rescaled PNDs $k_i^{\rm can}=u^{\rm can}+\hat\nu_i^{\rm can}$ have then unit spatial vectors
\bea
\label{hatnuatypeI}
\hat\nu_1^{\rm can}&=&\sin\alpha\cos\beta e_1^{\rm can}+\sin\alpha\sin\beta e_2^{\rm can}+\cos\alpha e_3^{\rm can}
=\hat\nu_1^{\rm can}(\alpha,\beta)
\,,\nonumber\\
\hat\nu_2^{\rm can}&=&
\hat\nu_1^{\rm can}(\pi-\alpha,-\beta)
\,,\nonumber\\
\hat\nu_3^{\rm can}&=&
\hat\nu_1^{\rm can}(\pi-\alpha,\pi-\beta)
\,,\nonumber\\
\hat\nu_4^{\rm can}&=&
\hat\nu_1^{\rm can}(\alpha,\pi+\beta)
\,,
\eea
with
\beq
\beta={\rm arctan}\left(\frac{2{\rm Re}(\lambda_1)}{1-|\lambda_1|^2}\right)\,,\qquad
\alpha={\rm arccos}\left(\frac{2{\rm Im}(\lambda_1)}{1+|\lambda_1|^2}\right)\,.
\eeq
The quadratic invariant \eqref{Omega1234quad} is then given by
\beq
\label{Omega1234quad2}
\frac{1}{2!3!4!}[\Omega_{1234}^{\rm can}]_{\alpha\beta\gamma\delta}[\Omega_{1234}^{\rm can}]^{\alpha\beta\gamma\delta}=-\frac{1024(1-|\lambda_1|^2)^2}{3(1+|\lambda_1|^2)^6}[{\rm Re}(\lambda_1)]^2[{\rm Im}(\lambda_1)]^2\,.
\eeq

\subsubsection{Petrov spacetime}

The Petrov spacetime \cite{Petrov1962} is a homogeneous vacuum solution with line element given by
\beq
k^2 ds^2= e^x [\cos (\sqrt{3}x)(dt^2-dz^2)+2\sin (\sqrt{3}x) dt dz]-dx^2-e^{-2x}dy^2\,,
\eeq
where $k>0$ is a constant parameter and $0<\sqrt{3}x<\pi/2$.
The orthonormal frame associated with the principal NP frame is given by
\bea
e_0^{\rm can}&=& ke^{-x/2}\left[\cos \left(\frac{\sqrt{3}x}{2}\right)\partial_t+\sin \left(\frac{\sqrt{3}x}{2}\right)\partial_z\right]
\,,\nonumber\\
e_1^{\rm can}&=& \frac{k}{\sqrt{2}}(\partial_x-e^{x}\partial_y)
\,,\nonumber\\
e_2^{\rm can}&=& \frac{k}{\sqrt{2}}(\partial_x+e^{x}\partial_y)
\,,\nonumber\\
e_3^{\rm can}&=& ke^{-x/2}\left[-\sin \left(\frac{\sqrt{3}x}{2}\right)\partial_t+\cos \left(\frac{\sqrt{3}x}{2}\right)\partial_z\right]
\,,
\eea
leading to the following nonvanishing Weyl scalars
\bea
\psi_0&=&\psi_4=-\frac{k^2\sqrt{3}}{2}e^{i\pi/6}
\,,\nonumber\\
\psi_2&=& -\frac{k^2}{2}e^{-i\pi/3}=-k^2-\psi_4\,.
\eea

The four linear independent PNDs $k_i^{\rm can}=u^{\rm can}+\hat\nu_i^{\rm can}$ have unit spatial vectors \eqref{hatnuatypeI} with 
\beq
\lambda_1=e^{-i\pi/3}+e^{-i\pi/6}
=\frac12(1-i)(1+\sqrt{3})\,,
\eeq
so that $\beta=-\pi/4$ and $\alpha={\rm arccos}(-\sqrt{3}/3)$.
The value of the quadratic invariant \eqref{Omega1234quad2} is $-\frac{64}{81}$.
The contraction of $\Omega_{1234}^{\rm can}$ with the Weyl tensor is identically vanishing.

\subsubsection{Dunn and Tupper spacetime}

The Dunn and Tupper solution~\cite{DT1} represents a spatially homogeneous spacetime with line element
\begin{equation}
ds^2=dt^2-\frac{t^2}{(m-n)^2}dx^2-t^{-2(m+n)}(e^{-2x}dy^2+e^{2x}dz^2)\,,
\label{(3.4)}
\end{equation}
where  $m\not=n$ are two constant parameters.
The source is a perfect fluid with 4-velocity $U=\partial_t$, and energy density and pressure given by
\begin{equation}
\rho= \frac{m^2+mn+n^2}{t^2} \,,\qquad
p = -\frac{4mn}{t^2}\,,\qquad
mn\le 0\,,
\label{(3.6)}
\end{equation}
respectively, provided that $m$ and $n$ satisfy the additional constraint $m(2m+1)+n(2n+1)=0$.

A spatial triad adapted to the observer $u=\partial_t\equiv e_0$ is given by
\beq
\label{frame_u}
e_1=\frac{m-n}{t}\partial_x\,,\quad
e_2=e^x t^{m+n}\partial_y\,,\quad
e_3=e^{-x} t^{m+n}\partial_z\,,
\eeq
with associated transverse NP frame with nonvanishing Weyl scalars 
\begin{eqnarray}
\psi_0&=&-\psi_4=-\frac{(m+n+1)(m-n)}{t^2}\,,
\nonumber\\
\psi_2&=& -\frac{(m-n)^2}{t^2}\,,
\end{eqnarray}
so that the spacetime is generally of Petrov type I.
It becomes of Petrov type D in the special case $m=-n-1$.

A canonical NP frame (with NP frame vectors $n$ and $m$ not to be confused with the spacetime parameters denoted by the same letters) is obtained by performing a type III null rotation $l\to {\mathcal A}^{-1}l\,,\quad n\to {\mathcal A}n \,,\quad  m \to e^{i\theta}m\,,\qquad \bar m\to e^{-i\theta}\bar m$, which leaves $\psi_2$ unchanged, whereas $\psi_0\to {\mathcal A}^{-2}e^{2i\theta}\psi_0$ and $\psi_4 \to {\mathcal A}^2e^{-2i\theta} \psi_4$, with $\theta=\frac{\pi}{4}$ and ${\mathcal A}^2=\sqrt{-\psi_0/\psi_4}=1$.
With respect to that frame the four linear independent PNDs $k_i^{\rm can}=u^{\rm can}+\hat\nu_i^{\rm can}$ have unit spatial vectors \eqref{hatnuatypeI} with 
\beq
\lambda_1=e^{-i\pi/4}\sqrt{3\xi+\sqrt{9\xi^2+1}}\,,
\eeq
where
\beq
\xi=\frac{|m-n|}{3|m+n+1|}\,.
\eeq
The value of the quadratic invariant \eqref{Omega1234quad2} turns out to be $-192\xi^2/(1+\sqrt{9\xi^2+1})^4$.
The contraction of $\Omega_{1234}$ with the Weyl tensor is identically vanishing.

In the limiting type D case, $m=-n-1$, one finds that $\Omega_{12}=2l\wedge n=-2e_0\wedge e_1$ with dual ${}^*\Omega_{12}=-2e_2\wedge e_3$, and its quadratic invariant $\frac12[\Omega_{12}]^2$ reduces to the constant value $-4$. 
Finally, $\Omega_{12}$ satisfies Maxwell-like equations with current $J=\frac{4}{t}e_1$, so that the rescaled quantity $t^{-2}\Omega_{12}$ is divergence-free.

\section{Concluding remarks}

We have studied some algebraic and differential properties of the wedge products of $p$ distinct PNDs associated with the eigen-bivectors of the Weyl tensor in a number of explicit exact solutions of the Einstein field equations, systematically analyzing them with respect to convenient observer families linked naturally to the PNDs in each case, after developing the necessary tools for generic spacetimes.
Besides the well-known vacuum metrics of Kerr and Kasner, these spacetime examples include several poorly known spacetimes like the Robinson-Trautman scalar field solution, the  Dunn-Tupper, Bonnor-Davidson and Allnutt perfect fluid solutions and the Petrov vacuum solution.
These concrete examples help shed new light on hidden relationships between the Petrov eigenvalues $\lambda_i$ and the wedge products of the PNDs, while taking into account the role of the otherwise arbitrary normalization factors in a convenient rescaling of the PNDs. 

For any pair of distinct PNDs, one can introduce a center of velocity frame in which the observer sees those null vectors to have opposite directions in its local rest space. These are useful for simplifying the expressions for the various scalars evaluated by that observer, since the magnetic part of the PND wedge products always vanishes identically.
For example, in the case of two distinct PNDs (type D and III) the canonical frame is also a center of velocity frame, and the quadratic invariant associated with the wedge of the PNDs has constant value $-4$ as a general result.
In the Kerr spacetime the Carter observers play the role of the center of velocity observers, and the static observers and ZAMOs are boosted in the $t$-$\phi$ plane with respect to them. Decomposing the PNDs along these new observers thus implies a rescaling by the corresponding $\gamma$-factors, so that their wedge products and quadratic invariants appear rescaled by a factor of $\gamma^2$ and $\gamma^4$, respectively. 
For the case of more than two distinct PNDs (type I and II), there exist center of velocity frames adapted to each pair of independent PNDs.

\section*{Acknowledgments}

D.B. acknowledges sponsorship of the Italian Gruppo Nazionale per la Fisica Matematica (GNFM) of the Istituto Nazionale di Alta Matematica (INDAM).


\end{document}